\documentclass[prd, preprint, longbibliography, 12pt]{revtex4-1}
\usepackage{amsmath}
\usepackage{amssymb}
\usepackage{setspace}
\usepackage{graphicx}
\usepackage{natbib}
\usepackage{float}
\usepackage{tensor}
\usepackage[utf8]{inputenc}
\usepackage{amsfonts}
\usepackage{braket}
\usepackage{esint}
\usepackage{breqn}
\usepackage{IEEEtrantools}

\begin{document}
 
 % ! TEX spellcheck
 %
%\ \vskip 1.0 in

\begin{center}
{ \large \bf Quantum gravity, minimum length, and holography}
%\large \it - Karolyhazy uncertainty relation, Planck scale foam, and  holography - \\

%\smallskip

\vskip 0.2 in

{\large{\bf Tejinder P.  Singh }}

\medskip

%{\it $^a${UM-DAE Centre for Excellence in Basic Sciences, Mumbai, 400098, India}\\

{\it Tata Institute of Fundamental Research,}
{\it Homi Bhabha Road, Mumbai 400005, India}\\
\bigskip
  { \tt e-mail: tpsingh@tifr.res.in}

\end{center}

\centerline{\bf ABSTRACT}
\noindent  The Karolyhazy uncertainty relation is the statement that if a device is used to measure a length $l$, there will be a minimum uncertainty $\delta l$ in the measurement, given by $(\delta l)^3 \sim  L_P^2\; l$. This is a consequence of combining the principles of quantum mechanics and general relativity. In this Letter we show how this relation arises in our approach to quantum gravity, in a bottom-up fashion, from the matrix dynamics of atoms of space-time-matter. We use this relation to define a space-time-matter foam at the Planck scale, and to argue that our theory is holographic. By coarse graining over time scales larger than Planck time, one obtains the laws of quantum gravity. Quantum gravity is not a Planck scale phenomenon; rather it comes into play whenever no classical space-time background is available to describe a quantum system. Space-time and classical general relativity arise from spontaneous localisation in a highly entangled quantum gravitational system. The Karolyhazy relation continues to hold in the emergent theory. An experimental confirmation of this relation will constitute a definitive test of the quantum nature of gravity.
\bigskip

\bigskip
The Karolyhazy uncertainty relation  \cite{Karolyhazi:66} is the statement that if one uses a measuring device to measure a length $l$, there will be a minimum uncertainty $\delta l$ in this measurement, given by the relation
\begin{equation}
(\delta l)^3 \sim L_{P}^2 \; l
\end{equation}
where $L_P$ is Planck length. This relation is a consequence of taking into account the intrinsic quantum mechanical uncertainty present in any measuring device, as well as of the fact that the device will have a gravitational field associated with itself. The quantum uncertainty can be reduced by making the device more massive, but increasing the mass too much makes the device into a black hole. The optimal choice of mass which balances quantum uncertainty and gravity results in the above constraint on the accuracy of the length measurement.

 The Karolyhazy relation has been studied by several authors \cite{Karolyhazi:86, RMP:2012, Ng:1994, Ng:1999, Sasakura:1999}, and has been derived in more than one way. For a nice recent review, see Ng \cite{Ng:2019}. This same review also explains the relevance of this relation to Wheeler's Planck scale space-time foam, and its implication that information is stored holographically, in a quantum gravitational system. Given that this relation arises in a very general way by combining the principles of quantum mechanics and general relativity, it can be regarded as a prediction of quantum gravity, and its experimental confirmation will be a proof for the quantum nature of gravity \cite{Amelio:1999, Ng:2001}. There is an analogous uncertainty relation for measurement of a time-interval $T$, given by
\begin{equation}
(\delta T)^3  \sim \tau_P^2 \; T
\label{untime}
\end{equation}
where $\tau_{P}$ is Planck time \cite{Ng:2019}.

In the present note, we explain how the Karolyhazy relation, and its consequent implications (Planck scale foam, holography) arise in our recently proposed quantum theory of gravity \cite{maithresh2019}. As we will see below, this relation is already present at the Level 0 matrix dynamics of our theory (for a description of the levels, see Fig. 1 below). Thus it arises in a bottom-up fashion; not by combining the principles of quantum mechanics and general relativity. These two theories are emergent in our approach, and emerge from the underlying matrix dynamics. In fact, as we shall see, the emergence of the two theories is precipitated by the Karolyhazy relation. It is also striking that the Karolyhazy relation depends on Newton's constant $G$ and Planck's constant $\hbar$ only through $L_P^2$. This is consistent with our quantum gravity theory, where the only fundamental constants at Level 0 are Planck length and Planck time; whereas $G$ and $\hbar$ emerge only subsequently, at Level I. We next give an overview of the new theory, before moving on to discuss the length uncertainty relation.

{
{\it Overview of the new matrix dynamics: Spontaneous Quantum Gravity / The Aikyon theory}:  The starting point for the development of this theory is quantum foundational. We seek an equivalent reformulation of quantum [field] theory which does not depend on classical spacetime. To do so, we first employ Adler's theory of trace dynamics, which is a Lagrangian dynamics in which classical c-number degrees of freedom are raised to the status of matrices/operators. These matrix valued dynamical variables obey the Euler-Lagrange [equivalently, Hamilton's] equations of motion, and the theory is assumed to hold at the Planck scale. Space-time however is assumed to be flat Minkowski space-time, and gravity is not incorporated in trace dynamics. The theory possesses a novel conserved charge, it being the sum of the commutators $[q_i, p_i]$ over all dynamical degrees of freedom. When the theory is coarse-grained over time intervals much larger than Planck time, this charge gets equipartitioned over all degrees of freedom and each commutator is set equal to $i\hbar$. Quantum theory emerges from the underlying trace dynamics as a consequence of this coarse-graining. We generalised trace dynamics to incorporate gravity into the theory, by raising space-time points also to the status of matrices / operators, and bringing in a new time parameter, the Connes time, to describe evolution. This is Level 0. The sought for quantum theory without classical spacetime is obtained from this generalised trace dynamics by coarse-graining the underlying theory over many Planck time / length scales. This is Level I.  If a large number of degrees of freedom get entangled, spontaneous localisation results, and classical space-time geometry and macroscopic bodies emerge, obeying the laws of general relativity. This is Level III. Those quantum degrees of freedom which have not undergone spontaneous localisation, can ether be described by the laws of generalised trace dynamics, or by those of quantum field theory on the emergent background classical space-time. This is Level II. It turns out that in order to define spin in this theory, one has to introduce the internal gauge symmetries of the standard model, and double the number of 
space-time dimensions to eight, leading to an 8-D non-commutative physical space. This is the space of octonions, and we have recently proposed a theory of unification in this space
\cite{Adler:04, Singhdice, Singhspin, MPSingh, Singh2020DA}. }

The fundamental building blocks of the matrix dynamics at Level 0 are the atoms of space-time-matter, each of which is described by the following action principle \cite{maithresh2019}
 \begin{equation}
\frac{L_P}{c} \frac{S}{C_0}  =  \frac{1}{2} \int d\tau \; Tr \bigg[\frac{L_P^2}{L^2c^2}\; (\dot{q}_B +\beta_1 \dot{q}_F)\;(\dot{q}_B +\beta_2 \dot{q}_F) \bigg]
\label{stmaction}
\end{equation}
For an explanation of the notation, the reader is referred to \cite{maithresh2019}. The universe is made of enormously many STM atoms, and labelling the action of the $i$-th atom as $S_i$, the total action of the (gravitation + Dirac fermions) of the universe is $S_{total}=\sum_i S_i$:
 \begin{equation}
\frac{L_P}{c} \frac{S_{total} }{C_0}  =  \frac{1}{2} \int d\tau \; \sum_i\; Tr \bigg[\frac{L_P^2}{L^2c^2}\; (\dot{q}_B +\beta_1 \dot{q}_F)\;(\dot{q}_B +\beta_2 \dot{q}_F) \bigg]_i
\label{atot}
\end{equation}
In principle, this action contains everything needed to describe the quantum gravitational universe, and the emergent quantum (field) theory, as well as the classical space-time geometry described by the laws of general relativity. 

By varying the action (\ref{stmaction}) of the STM atom with respect to the bosonic operator $q_B$ and the fermionic operator $q_F$, and after integrating once the resulting equations of motion, the following first integrals result \cite{maithresh2019}
\begin{align}
    2\dot{q}_B +(\beta_1+\beta_2)\dot{q}_F = c_1 = \frac{2}{a}\; p_B \\
    \dot{q}_B(\beta_1+\beta_2)+ \beta_1 \dot{q}_F \beta_2 +\beta_2 \dot{q}_F \beta_1 = c_2 =\frac{2}{a}\; p_F
    \label{firstint}
\end{align}
where $c_1$ and $c_2$ are constant bosonic and fermionic matrices, respectively, being the values of the bosonic momentum and the fermionic momentum. Here, $a=L_P^2/L^2c^2$. We defined the generalised Dirac operator $D$ in terms of its bosonic and fermionic compoinents ($D_B$ and $D_F$ respectively) 
\begin{equation}
\frac{1}{Lc}\;  \frac{dq}{d\tau}\sim D \equiv D_B + D_F ; \qquad D_B \equiv \frac{1}{Lc}\;  \frac{dq_B}{d\tau} ; \qquad   D_F \equiv \frac{\beta_1 + \beta_2}{2Lc} \frac{dq_F}{d\tau}
\end{equation} 
Hence the first of the two integrals of motion above can be written as an eigenvalue equation
\begin{equation}
[D_B + D_F] \psi = \lambda \psi \equiv (\lambda_R + i \lambda_I)\psi \equiv \bigg(\frac{1}{L} + i \frac{1}{L_I}\bigg)\psi
\label{gen}
\end{equation}
where the eigenvalues $\lambda$, assumed to be $c$-numbers, are independent of the Connes time $\tau$, and $\lambda_R$ and $\lambda_I$ are its real and imaginary parts respectively. 

It is this generalisation of Dirac equation which is of interest to us here, as regards the Karolyhazy uncertainty relation. As explained in \cite{maithresh2019}, it is the relative magnitude of the real and imaginary parts of the eigenvalue, $\lambda_R$ and $\lambda_I$, which decide upon the properties of the STM atom. If $L\ll L_I$, the STM atom is quantum in nature, behaving like a Dirac fermion, and obeying the Level 0 analogue of the usual Dirac equation. If, on the other hand, $L\gg L_I$, spontaneous localisation ensues, resulting in classical behaviour. Einstein-Hilbert action with point-like matter sources is recovered, provided we choose $L_I = L^3 / L_p^2$. In \cite{maithresh2019}, we assumed this relation, but now we can ensure that this relation already arises from the action at Level 0, by assuming that the two $\beta$ matrices scale as $L_p^2/L^2$. In the classical limit, this relation between $L$ and $L_I$ implies that $L$, which was previously identified with Compton wavelength, is much smaller than Planck length. This in turn is consistent with the STM atom behaving like a black hole [the Schwarzschild radius $L_P^2 / L$ far exceeds both $L_P$ as well as $L$].

However, if we  do not ignore either the classical aspect (i.e. $L_I$) or the quantum aspect (i.e. $L$), we can consider 
how one term corrects the magnitude of the  other. In the classical situation $L_I \ll L$, both $L$ and $L_I$ are much smaller than Planck length, and this case is hence not of much interest. On the contrary, in the quantum situation $L_I\gg L$, both $L_I$ and $L$ are much greater than Planck length, and we can think of the $1/L_I$ term as providing a small correction to $1/L$. In other words, $L_P^2/L^3$ is a small correction to $1/L$, or  
$L$ is a small correction to $L_I$. We can hence identify $L$ with the $\delta l$ of the Karolyhazy relation, and $L_I$ with $l$. It then immediately follows from the above relation $L_I = L^3 / L_p^2$ between $L_I$ and $L$, that
\begin{equation}
(\delta l)^3 \sim L_{P}^2 \; l
\end{equation}
This of course is the same as the Karolyhazy relation, but now arrived at as a property of the STM atom. If we are to treat the STM atom as a quantum object of length $L=\hbar/mc$, this length is a small correction $\delta l=L$ to the associated  length $l=L_I =\hbar^2/Gm^3$, which happens to be the gravitational decoherence length.  Thus, this is also the same as the statement $m\ll m_{Pl}$. The Karolyhazy relation is now an intrinsic property of an atom of space-time-matter at Level 0. We do not have to separately talk of a measuring device, the quantum uncertainty principle, or Schwarzschild radius of the device - these are all emergent concepts of Level II and III; and the Karolyhazy relation is more fundamental than Level II and Level III concepts.
It is easy to check that the correction $L_I$ to $L$ amounts, from the point of view of the Dirac equation, to modifying the mass term $mc/\hbar$ to an effective mass $m_{eff}$ including  an imaginary correction term:
\begin{equation}
m_{eff} = \frac{mc}{\hbar}\; \bigg[ 1+i\frac{m^2}{m_{Pl}^2}\bigg]
\end{equation}

If in Eqn. (\ref{gen}) we consider the special case of an STM atom for which $L_I = L_P$, then of course from $L_I = L^3 / L_P^2$ we get $L=L_P$. This represents Planck scale `space-time-matter foam'. It can be understood as the STM atoms at the Planck scale, undergoing extremely rapid (on the scale of Planck time, in Connes time $\tau$) quantum expansion (indicated by $L=L_P$) and spontaneous localisation / contraction (indicated by $L_I = L_P$). Thus we have introduced the concept of the Planck scale foam in a bottom-up fashion. It is not that Planck scale space-time foam is defined by quantizing a pre-existing classical space-time geometry. Rather, classical space-time geometry emerges from coarse graining (over scales larger than Planck scale) the matrix dynamics of the STM atoms. The STM foam is defined in terms of the matrix dynamics at Level 0. In the language of the Karolyhazy relation the foam is given by $l=\delta l = L_P$, as has already been noted and discussed earlier by Ng \cite{Ng:2019} and other researchers. The difference between earlier discussions and ours is that for us the foam is of space-time-matter, not of space-time. At the Planck scale, there is no distinction between space-time and matter.

We take this opportunity to emphasize that there is an imaginary part to the eigenvalues of the generalised Dirac operator, precisely because the Hamiltonian of the theory is not self-adjoint, at Level 0. There is no reason why the fundamental Hamiltonian of the universe should be self-adjoint at the Planck scale. We only require that the emergent Hamiltonian, after coarse graining over scales larger than Planck scales, should be self-adjoint, in quantum field theory, and in classical dynamics. The presence of an anti-self-adjoint part in the fundamental Hamiltonian (this part arises naturally from the structure, and is not added by hand in an ad hoc way) enables the classical limit (absence of macroscopic superpositions in matter and in space-time geometries) to arise dynamically. This happens by the process of spontaneous localisation. One does not have to appeal to one or the other interpretations of quantum theory to explain the nature of the quantum-to-classical transition.

We also note that $L_I$ does not scale linearly with $L$, but faster, as $L^3$. This is what makes quantum effects relatively less important, compared to classicality, as one moves to $L$ values larger than Planck length. If $L_I$ were to scale linearly with $L$, then on all scales quantum effects would remain as important as classical effects, and classical emergence would not be possible.

Furthermore, the Karolyhazy relation implies that our theory is holographic: the amount of quantum information in a region of size $L_I$ grows as the surface area $L_I^2$ of its boundary, not as its volume $L_I^3$. Given that $L$ is a measure of minimum quantum uncertainty, we 
can assume that a length $L_I$ can be divided into $L_I/L$ discrete cells, but no more. Thus, a measure of the number of quantum information units in a volume $L_I^3$ is $(L_I/L)^3$ which by virtue of the Karolyhazy relation is equal to $L_I^2/L_P^2$ thus implying that our theory is holographic. Such a conclusion - namely that a quantum gravity theory obeying the Karolyhazy relation is holographic - was already arrived at earlier by Ng \cite{Ng:2019}, and possibly also by other researchers. We have only reiterated it in the context of our theory. Note that holography results because the uncertainty volume grows with $L_I$. Were it to be a constant (say if it was $L_P^3$) then quantum information would increase as the volume $L_I^3$, not as the area $L_I^2$. As argued by Ng, this holography is also the reason why the black hole entropy grows as its area. It is noteworthy that this result for the entropy, recently derived rigorously by us in our theory \cite{maithresh2019b}, is also a direct consequence of the Karolyhazy relation.

We now outline, building on our earlier discussion, how our Karolyhazy relation can be mapped to the more conventional one, which derives from combining the principles of general relativity and quantum mechanics. We note that the conventional derivation is ad hoc /semi-classical in nature, because the gravitational field of the measuring device is treated classically. Our bottom-up derivation is rigorous. Nonetheless, it will be useful to relate the rigorous derivation to the semi-classical one. For this, we build on our earlier analysis to demonstrate how one goes from the matrix dynamics of Level 0 to classical general relativity at Level III and quantum (field) theory at Level II.

To begin with, we recall how one gets from Level 0 to Level I, where Level I is quantum gravity. We assume that we are not interested in examining microscopically the Level 0 dynamics of the STM atoms, taking place on the Planck scale.This dynamics plays out in the Hilbert space of Level 0, marked by evolution in Connes time $\tau$. On the Planck time scale, we encounter space-time-matter foam, as described above - quantum evolution and spontaneous collapse of STM atoms resulting in generation of Planck scale curvature. Suppose we are not interested in dynamics at this level of detail - then we must coarse grain over time scales much larger than Planck time $\tau_{P}$. And we must ask what is this coarse-grained dynamics, resulting from the underlying matrix dynamics? To answer this question, we employ the methods of statistical mechanics, following Adler's scheme in trace dynamics \cite{Adler:04}. We examine the dynamics of a large collection of STM atoms in phase space, and assume that the long time average [i.e. times longer than Planck time] can be given by the ensemble average [i.e. the ergodic hypothesis holds]. The ensemble average is found by using the methods of statistical mechanics to maximise the von Neumann entropy made from the probability distribution in the matrix dynamics phase space. The equilibrium distribution then describes the coarse grained mean dynamics of an STM atom, on time scales larger than Planck time.

We ask as to what kind of interaction between the STM atoms drives their ensemble to statistical equilibrium? Tentatively, we introduce here the concept of `collision' of two STM atoms in Hilbert  space. Two STM atoms $q_1 = q_{1B} + q_{1F}$ and $q_2 = q_{2B}+ q_{2F}$ will be said to collide in Hilbert space at Connes time $\tau$ if $q_{1}(\tau) = q_2(\tau)$, i.e. $q_{1B}(\tau)=q_{2B}(\tau)$ and 
$q_{1F}(\tau)=q_{2F}(\tau)$.

The collision will be said to be instantaneous (a delta-function interaction at time $\tau$) and elastic if after the collision the two STM atoms bounce off each other, while obeying conservation of momentum ($p_F$ and $p_B$ separately), trace Hamiltonian, and the Adler-Millard charge. Interestingly, a consistent system of equations can be set up to describe such a collision, as we now show.

As we saw above, in Eqns. (5) and (\ref{firstint}), the first integrals give the expressions for the constant bosonic and fermionic momentum.  These equations can be integrated to solve for $q_B(\tau)$ and $q_F(\tau)$, these describe the trajectory of the STM atom  in Hilbert space:
\begin{align}
     \dot{q}_B &= \frac{1}{2}\bigg[c_1 -(\beta_1+\beta_2)(\beta_1-\beta_2)^{-1}\big[2c_2 -c_1(\beta_1+\beta_2) \big](\beta_2-\beta_1)^{-1} \bigg] \label{qb} \\
     \dot{q}_F &= (\beta_1-\beta_2)^{-1}\big[2c_2-c_1(\beta_1+\beta_2)\big](\beta_2-\beta_1)^{-1} \label{qf}
\end{align}
This means that the velocities $\dot{q}_B$ and $\dot{q}_F$ are constant,  and $q_B$ and $q_F$ evolve linearly in Connes time. The trace Hamiltonian is given by
\begin{equation}
    \textbf{H} = \text{Tr} \frac{2}{a} \bigg[(p_B\beta_1-p_F)(\beta_2-\beta_1)^{-1}(p_B\beta_2-p_F)(\beta_1-\beta_2)^{-1}
    \bigg]
\end{equation}
and the Adler-Millard charge by the commutators
\begin{equation}
\tilde{C} = [q_B, p_B] - \{q_F, p_F\}
\end{equation}
If $\tau=\tau_c$ is the time at which collision occurs, that fixes the values of $q_B$ and $q_F$ at the time of the collision. And the constant values  for the trace Hamiltonian and the Adler-Millard charge are known in terms of the constant values of the momenta
$p_B$ and $p_F$. This allows us to set up a two body elastic collision problem, as in Newtonian mechanics, to determine the final momenta $p_{1B}'$, $p_{1F}'$, $p_{2B}'$ and $p_{2F}'$, in terms of their initial momenta. The four final momenta can be determined from the following four equations, applying conservation of the two momenta, the trace Hamiltonian, and the Adler-Millard charge:
\begin{equation}
p_{1B} + p_{2B} = p_{1B}' + p_{2B}'
\end{equation}
 \begin{equation}
p_{1F} + p_{2F} = p_{1F}' + p_{2F}'
\end{equation}
\begin{equation}
{\bf H_1} + {\bf H_2} = {\bf H_1}' + {\bf H_2}'
\end{equation}
\begin{equation}
\tilde{C}_1 + \tilde{C}_2 = \tilde{C}_1' + \tilde{C}_2 '
\end{equation}
We would like to suggest that such collisions between STM atoms drive the system of many STM atoms to equilibrium in phase space, over time intervals much larger than Planck time. As shown in trace dynamics, equilibrium corresponds to the equipartition of the Adler-Millard charge, and the equipartitioned value is identified with Planck's constant $\hbar$. Furthermore, Adler showed in trace dynamics, and the same holds for our theory, that the mean dynamics at equilibrium is described by quantum commutators, and Heisenberg equations of motion:
\begin{equation}
[q_B, p_B] = i \hbar; \qquad \{q_{FS}, p^f_{FAS}\} = i\hbar; \qquad \{q_{FAS}, p^f_{FS}\} = i\hbar
\end{equation}
\begin{equation} 
i\hbar \frac{\partial q_B}{\partial \tau} = [q_B, { H}_S]; \qquad i\hbar \frac{\partial p_B}{\partial \tau} =  [p_B, { H}_S]; \qquad i\hbar \frac{\partial q_F}{\partial \tau} = [q_F,{ H}_S]; \qquad i\hbar \frac{\partial p^f_F}{\partial \tau} =  [p^f_F, { H}_S]
\end{equation}
In analogy with quantum field theory, one can transform from the above Heisenberg picture, and write a Schr\"{o}dinger equation for the wave-function $\Psi(\tau)$ of the full system:
\begin{equation}
i\hbar \; \frac{\partial \Psi}{\partial \tau} = { H_{Stot}} \Psi (\tau)
\end{equation}
where ${ H_{Stot}}$ is the sum of the self-adjoint parts of the Hamiltonians of the individual STM atoms.

This emergent theory at Level I is quantum gravity. It is quantum gravity in the sense that there are quantum commutation relations and Heisenberg equations of motion for both the gravity sector $q_B$ and the matter sector $q_F$. But there is no classical space-time yet; evolution is still with respect to Connes time $\tau$. This brings us to the following significant question: having coarse-grained over the Planck scale, how do we still get quantum gravity? The answer is that quantum gravity is {\it not} a Planck scale phenomenon! Our Planck scale theory is the Level 0 matrix dynamics of STM atoms. On the other hand, quantum gravity is how we should describe the dynamics of quantum systems when there is no background classical space-time available. In fact, this is how quantum systems are correctly described - their gravity part, as well as the matter part, both obeys quantum laws. If we ignore their gravity part, and instead use the Level III classical geometry to describe space-time, we obtain the Level II quantum (field) theory on a classical space-time, which is how we conventionally study quantum theory. But this is an approximate description of a quantum system. It seems entirely possible that when we take into account the quantum nature of the $q_B$ associated with a quantum field, we will see possible quantum gravitational modifications to quantum theory, and this need not be just a Planck scale effect, and could perhaps be detected at lower energies. For instance, if we were to ask how to describe the quantum gravitational field of an electron, it will be through the equations of Level I given above.

Because the Schr\"{o}dinger picture is available at Level I, it becomes possible to also talk of the entanglement of states of STM atoms. Entanglement does not require classical space-time; in fact it is well-known that quantum entanglement is oblivious to space-time. Thus it is very reasonable to say that entanglement is a Level I phenomenon, where evolution in Connes time is available, but there is no space-time. Thus we no longer talk of how far apart the entangled particles are, or whether Alice affected Bob, or Bob affected Alice. Entanglement being a Level I phenomenon, does not respect the rules of distance or causality; the latter are defined only at the emergent levels II and III \cite{Singh2019qg}. 

As demonstrated in some detail in \cite{maithresh2019}, the above quantum gravity theory arises at statistical equilibrium at Level I. The dynamics is determined by the ensemble average of the self-adjoint part of the trace Hamiltonian at Level 0. The anti-self-adjoint part of the Hamiltonian provides corrections to the equilibrium theory; and these are represented as stochastic fluctuations:
\begin{equation}
i\hbar \; \frac{\partial \Psi}{\partial \tau} = [{ H_{Stot}} + {\cal H}(\tau)] \Psi (\tau)
\end{equation}
These fluctuations, because they have an anti-self-adjoint part, cause breakdown of quantum superposition - what is generally referred to as spontaneous localisation. This spontaneous localisation is extremely rare and infrequent for a single STM atom. However, if we consider an entangled state of many STM atoms, spontaneous localisation becomes more and more frequent (the amplification mechanism). Until, if the number of entangled STM atoms becomes macroscopically large, the localisation process becomes extremely rapid. This leads to the emergence of space-time and classical material objects, obeying the laws of general relativity - this is the gravitational dynamics at the emergent Level III \cite{maithresh2019}. 

Thus the purely quantum (gravitational)  level is Level I, whereas the classical level is Level III. The quantum theory on a classical space-time that we practice is the hybrid Level II, as discussed in \cite{maithresh2019}. The conventional derivation of the Karolyhazy 
relation takes the classical (gravitational) behaviour of the measuring device from Level III, and its quantum behaviour from the hybrid Level II. This makes it quite clear that the conventional derivation is rather approximate, and somewhat heuristic in nature. What we have provided in this paper is the strictly rigorous derivation of the Karolyhazy relation, at Level 0. From there, one can relate to the conventional derivation, through the steps outlined above: Level 0 to Level I, and from there to Levels III and II.

An experimental confirmation of the Karolyhazy relation constitutes a definitive test of the quantum nature of gravity. We can list here some representative values of $\delta l = L_P^{2/3}\; l^{1/3}$ for a few different values of $l$:
\begin{equation}
\begin{split}
l= L_P,\; \delta l = L_P; \quad l = 10^{-13}\; {\rm cm}, \; \delta l = 10^{-26}\; {\rm cm}; \quad l= 1\; {\rm cm},\; \delta l = 10^{-22}\; {\rm cm};\\
l = 4 \; {\rm km}, \; \delta l = 10^{-20}\; {\rm cm}; \quad l=10^{6}\; {\rm km}, \; \delta l = 10^{-18}\; {\rm cm}; \quad l = 10^{28}\; {\rm cm}, \; \delta l = 10^{-13}\; {\rm cm}
\end{split}
\end{equation}
The strain  $\delta l /l$ decreases with increasing $l$ as $l^{-2/3}$. It is intriguing that when $l$ is of the order of the size of the universe, then $\delta l$ is of the order of the Compton wavelength of the proton. Various researchers have already earlier emphasized the importance of a laboratory test of the Karolyhazy uncertainty relation.
In parallel to testing this relation, another test would be the analogous one for time measurement, for which the uncertainty relation is given in (\ref{untime}) above. Another prediction of our theory is quantum interference / spontaneous localisation in time. Also, from the point of view of collapse models, in our theory the collapse time scale is likely given by $L_I/c = L^3/cL_P^2 = \hbar^2/Gcm^3$, if we take $L$ to be the Compton wavelength of the particle. The collapse rate hence grows as $m^3$, implying that ours is not a mass-proportional collapse model. However, for a nucleon we do get that the collapse rate is nearly the GRW value, $10^{-17}$\; s$^{-1}$. 

{
A dedicated experiment is currently being planned  \cite{s2020experiment} to test for quantum gravity induced length uncertainty relations of the type $(\Delta l)^3 \sim L_P^n \;  l^{3-n}$. of which the Karolyhazy relation is a special case. Also, from the above estimates we see that for $l$ of the order of a km, the uncertainty induces a strain $\delta l / l\sim 10^{-25}$ cm. This is a few orders of magnitude beyond Advanced LIGO sensitivity. A crucial piece of information missing from our present analysis is the noise spectrum of the stochasticity [predicted by the underlying theory], which leads to the Karolyhazy relation. Once one knows the noise spectrum, one could investigate further the feasibility of testing for this relation in a gravitational wave detector. }

In our opinion, this new approach to quantum gravity presents a fresh and original alternative to the problem, as represented in Fig. 1 below, borrowed from \cite{maithresh2019b}. 
\begin{figure}[!htb]
        \center{\includegraphics[width=\textwidth]
        {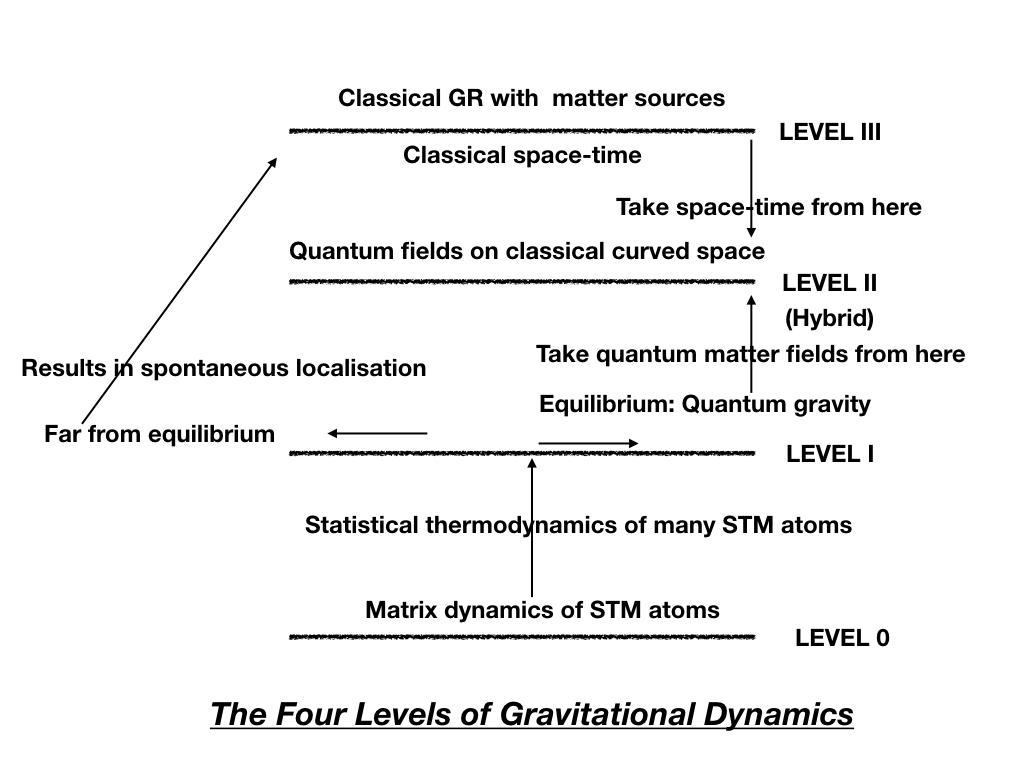}}
        \caption{\label{fig:my-label} The four levels of gravitational dynamics. In this bottom-up theory, the fundamental Level 0 describes the `classical' matrix dynamics of atoms of space-time-matter (STM). This level operates at the Planck scale. Statistical thermodynamics of these atoms brings us below Planck scale, to Level I: the emergent equilibrium theory is quantum gravity. Far from equilibrium, rapid spontaneous localisation results  in Level III: emergence of  classical space-time, obeying classical general relativity with matter sources. Level II is a hybrid level built by taking classical space-time from Level III and quantum matter fields from Level I, while neglecting the quantum gravitation of Level I. Strictly speaking, all quantum field dynamics takes place at Level I, but we approximate that to Level II. From \cite{maithresh2019b}.}
      \end{figure}
At the Planck scale, we do not have quantum gravity. Instead we have a matrix dynamics of atoms of space-time-matter, described by the action principle given in Eqn. (\ref{atot}) above. When the evolution of the STM atoms is averaged over time scales much larger than Planck scale, we get a quantum theory of gravity. This theory obeys the rules of quantum theory, both for the gravity sector, and for the matter sector. When a large number of STM atoms are entangled, rapid spontaneous localisation occurs, resulting in the emergence of classical space-time and its geometry, obeying the laws of classical general relativity with matter sources. If we now take matter to be quantum (i.e. from Level I) but gravity to be classical (i.e. from Level III) we get the usual quantum theory - it being an excellent approximation to the physics of Level I \cite{singh2019qf}.

\bigskip

\bigskip

I would like to thank Maithresh Palemkota, Hendrik Ulbricht and Daniel Goldwater for useful discussions, which have also contributed to some of the ideas in this paper.

\bigskip

\bigskip

\centerline{\bf REFERENCES}

\bibliography{biblioqmtstorsion}

%merlin.mbs apsrev4-1.bst 2010-07-25 4.21a (PWD, AO, DPC) hacked
%Control: key (0)
%Control: author (0) dotless jnrlst
%Control: editor formatted (1) identically to author
%Control: production of article title (0) allowed
%Control: page (1) range
%Control: year (0) verbatim
%Control: production of eprint (0) enabled
\def\polhk#1{\setbox0=\hbox{#1}{\ooalign{\hidewidth
  \lower1.5ex\hbox{`}\hidewidth\crcr\unhbox0}}} \def\cprime{$'$}
  \def\cprime{$'$}
\begin{thebibliography}{19}%
\makeatletter
\providecommand \@ifxundefined [1]{%
 \@ifx{#1\undefined}
}%
\providecommand \@ifnum [1]{%
 \ifnum #1\expandafter \@firstoftwo
 \else \expandafter \@secondoftwo
 \fi
}%
\providecommand \@ifx [1]{%
 \ifx #1\expandafter \@firstoftwo
 \else \expandafter \@secondoftwo
 \fi
}%
\providecommand \natexlab [1]{#1}%
\providecommand \enquote  [1]{``#1''}%
\providecommand \bibnamefont  [1]{#1}%
\providecommand \bibfnamefont [1]{#1}%
\providecommand \citenamefont [1]{#1}%
\providecommand \href@noop [0]{\@secondoftwo}%
\providecommand \href [0]{\begingroup \@sanitize@url \@href}%
\providecommand \@href[1]{\@@startlink{#1}\@@href}%
\providecommand \@@href[1]{\endgroup#1\@@endlink}%
\providecommand \@sanitize@url [0]{\catcode `\\12\catcode `\$12\catcode
  `\&12\catcode `\#12\catcode `\^12\catcode `\_12\catcode `\%12\relax}%
\providecommand \@@startlink[1]{}%
\providecommand \@@endlink[0]{}%
\providecommand \url  [0]{\begingroup\@sanitize@url \@url }%
\providecommand \@url [1]{\endgroup\@href {#1}{\urlprefix }}%
\providecommand \urlprefix  [0]{URL }%
\providecommand \Eprint [0]{\href }%
\providecommand \doibase [0]{http://dx.doi.org/}%
\providecommand \selectlanguage [0]{\@gobble}%
\providecommand \bibinfo  [0]{\@secondoftwo}%
\providecommand \bibfield  [0]{\@secondoftwo}%
\providecommand \translation [1]{[#1]}%
\providecommand \BibitemOpen [0]{}%
\providecommand \bibitemStop [0]{}%
\providecommand \bibitemNoStop [0]{.\EOS\space}%
\providecommand \EOS [0]{\spacefactor3000\relax}%
\providecommand \BibitemShut  [1]{\csname bibitem#1\endcsname}%
\let\auto@bib@innerbib\@empty
%</preamble>
\bibitem [{\citenamefont {Karolyhazy}(1966)}]{Karolyhazi:66}%
  \BibitemOpen
  \bibfield  {author} {\bibinfo {author} {\bibfnamefont {F.}~\bibnamefont
  {Karolyhazy}},\ }\bibfield  {title} {\enquote {\bibinfo {title} {Gravitation
  and quantum mechanics of macroscopic objects},}\ }\href@noop {} {\bibfield
  {journal} {\bibinfo  {journal} {Nuovo Cimento}\ }\textbf {\bibinfo {volume}
  {42A}},\ \bibinfo {pages} {390} (\bibinfo {year} {1966})}\BibitemShut
  {NoStop}%
\bibitem [{\citenamefont {Karolyhazy}\ \emph {et~al.}(1986)\citenamefont
  {Karolyhazy}, \citenamefont {Frenkel},\ and\ \citenamefont
  {Luk\'acs}}]{Karolyhazi:86}%
  \BibitemOpen
  \bibfield  {author} {\bibinfo {author} {\bibfnamefont {F.}~\bibnamefont
  {Karolyhazy}}, \bibinfo {author} {\bibfnamefont {A.}~\bibnamefont {Frenkel}},
  \ and\ \bibinfo {author} {\bibfnamefont {B.}~\bibnamefont {Luk\'acs}},\
  }\bibfield  {title} {\enquote {\bibinfo {title} {On the possible role of
  gravity in the reduction of the wave function},}\ }in\ \href@noop {} {\emph
  {\bibinfo {booktitle} {Quantum concepts in space and time}}},\ \bibinfo
  {editor} {edited by\ \bibinfo {editor} {\bibfnamefont {R.}~\bibnamefont
  {Penrose}}\ and\ \bibinfo {editor} {\bibfnamefont {C.~J.}\ \bibnamefont
  {Isham}}}\ (\bibinfo  {publisher} {Clarendon},\ \bibinfo {address} {Oxford},\
  \bibinfo {year} {1986})\BibitemShut {NoStop}%
\bibitem [{\citenamefont {Bassi}\ \emph {et~al.}(2013)\citenamefont {Bassi},
  \citenamefont {Lochan}, \citenamefont {Satin}, \citenamefont {Singh},\ and\
  \citenamefont {Ulbricht}}]{RMP:2012}%
  \BibitemOpen
  \bibfield  {author} {\bibinfo {author} {\bibfnamefont {Angelo}\ \bibnamefont
  {Bassi}}, \bibinfo {author} {\bibfnamefont {Kinjalk}\ \bibnamefont {Lochan}},
  \bibinfo {author} {\bibfnamefont {Seema}\ \bibnamefont {Satin}}, \bibinfo
  {author} {\bibfnamefont {Tejinder~P.}\ \bibnamefont {Singh}}, \ and\ \bibinfo
  {author} {\bibfnamefont {Hendrik}\ \bibnamefont {Ulbricht}},\ }\bibfield
  {title} {\enquote {\bibinfo {title} {Models of wave function collapse,
  underlying theories, and experimental tests},}\ }\href@noop {} {\bibfield
  {journal} {\bibinfo  {journal} {Rev. Mod. Phys.}\ }\textbf {\bibinfo {volume}
  {85}},\ \bibinfo {pages} {471 arXiv:1204.4325 [quant--ph]} (\bibinfo {year}
  {2013})}\BibitemShut {NoStop}%
\bibitem [{\citenamefont {Ng}\ and\ \citenamefont {van Dam}(1994)}]{Ng:1994}%
  \BibitemOpen
  \bibfield  {author} {\bibinfo {author} {\bibfnamefont {Y.~J.}\ \bibnamefont
  {Ng}}\ and\ \bibinfo {author} {\bibfnamefont {H.}~\bibnamefont {van Dam}},\
  }\bibfield  {title} {\enquote {\bibinfo {title} {Limit to space-time
  measurement},}\ }\href@noop {} {\bibfield  {journal} {\bibinfo  {journal}
  {Mod. Phy. Letts. A}\ }\textbf {\bibinfo {volume} {3}},\ \bibinfo {pages}
  {335} (\bibinfo {year} {1994})}\BibitemShut {NoStop}%
\bibitem [{\citenamefont {Ng}\ and\ \citenamefont {van Dam}(1995)}]{Ng:1999}%
  \BibitemOpen
  \bibfield  {author} {\bibinfo {author} {\bibfnamefont {Y.~J.}\ \bibnamefont
  {Ng}}\ and\ \bibinfo {author} {\bibfnamefont {H.}~\bibnamefont {van Dam}},\
  }\bibfield  {title} {\enquote {\bibinfo {title} {Remarks on gravitational
  sources},}\ }\href@noop {} {\bibfield  {journal} {\bibinfo  {journal} {Mod.
  Phy. Letts. A}\ }\textbf {\bibinfo {volume} {10}},\ \bibinfo {pages} {2801}
  (\bibinfo {year} {1995})}\BibitemShut {NoStop}%
\bibitem [{\citenamefont {Sasakura}(1999)}]{Sasakura:1999}%
  \BibitemOpen
  \bibfield  {author} {\bibinfo {author} {\bibfnamefont {N.}~\bibnamefont
  {Sasakura}},\ }\bibfield  {title} {\enquote {\bibinfo {title} {An uncertainty
  relation of space-time},}\ }\href@noop {} {\bibfield  {journal} {\bibinfo
  {journal} {Prog. Theor. Phys.}\ }\textbf {\bibinfo {volume} {102}},\ \bibinfo
  {pages} {169} (\bibinfo {year} {1999})}\BibitemShut {NoStop}%
\bibitem [{\citenamefont {Ng}(2019)}]{Ng:2019}%
  \BibitemOpen
  \bibfield  {author} {\bibinfo {author} {\bibfnamefont {Y.~J.}\ \bibnamefont
  {Ng}},\ }\bibfield  {title} {\enquote {\bibinfo {title} {Entropy and
  gravitation: From black hole computers to dark energy and dark matter},}\
  }\href@noop {} {\bibfield  {journal} {\bibinfo  {journal} {Entropy}\ }\textbf
  {\bibinfo {volume} {21}},\ \bibinfo {pages} {doi: 10.3390/e21111035
  arXiv:1910.00040 v1 [gr--qc]} (\bibinfo {year} {2019})}\BibitemShut {NoStop}%
\bibitem [{\citenamefont {Amelino-Camelia}(1999)}]{Amelio:1999}%
  \BibitemOpen
  \bibfield  {author} {\bibinfo {author} {\bibfnamefont {Giovanni}\
  \bibnamefont {Amelino-Camelia}},\ }\bibfield  {title} {\enquote {\bibinfo
  {title} {Gravity-wave interferometers as quantum gravity detectors},}\
  }\href@noop {} {\bibfield  {journal} {\bibinfo  {journal} {Nature}\ }\textbf
  {\bibinfo {volume} {398}},\ \bibinfo {pages} {216} (\bibinfo {year}
  {1999})}\BibitemShut {NoStop}%
\bibitem [{\citenamefont {Ng}(2001)}]{Ng:2001}%
  \BibitemOpen
  \bibfield  {author} {\bibinfo {author} {\bibfnamefont {Y.~J.}\ \bibnamefont
  {Ng}},\ }\bibfield  {title} {\enquote {\bibinfo {title} {From computation to
  black holes and space-time foam},}\ }\href@noop {} {\bibfield  {journal}
  {\bibinfo  {journal} {Phys. Rev. Lett.}\ }\textbf {\bibinfo {volume} {86}},\
  \bibinfo {pages} {2946} (\bibinfo {year} {2001})}\BibitemShut {NoStop}%
\bibitem [{\citenamefont {Palemkota}\ and\ \citenamefont {Singh}(2019
  DOI:10.1515/zna-2019-0267 arXiv:1908.04309)}]{maithresh2019}%
  \BibitemOpen
  \bibfield  {author} {\bibinfo {author} {\bibfnamefont {Maithresh}\
  \bibnamefont {Palemkota}}\ and\ \bibinfo {author} {\bibfnamefont
  {Tejinder~P.}\ \bibnamefont {Singh}},\ }\bibfield  {title} {\enquote
  {\bibinfo {title} {Proposal for a new quantum theory of gravity {III}:
  Equations for quantum gravity, and the origin of spontaneous localisation},}\
  }\href@noop {} {\bibfield  {journal} {\bibinfo  {journal} {Zeitschrift f\"ur
  Naturforschung A}\ }\textbf {\bibinfo {volume} {75}},\ \bibinfo {pages} {143}
  (\bibinfo {year} {2019 DOI:10.1515/zna-2019-0267
  arXiv:1908.04309})}\BibitemShut {NoStop}%
\bibitem [{\citenamefont {Adler}(2004)}]{Adler:04}%
  \BibitemOpen
  \bibfield  {author} {\bibinfo {author} {\bibfnamefont {Stephen~L.}\
  \bibnamefont {Adler}},\ }\href@noop {} {\emph {\bibinfo {title} {Quantum
  theory as an emergent phenomenon}}}\ (\bibinfo  {publisher} {Cambridge
  University Press, Cambridge},\ \bibinfo {year} {2004})\BibitemShut {NoStop}%
\bibitem [{\citenamefont {Singh}(2020{\natexlab{a}})}]{Singhdice}%
  \BibitemOpen
  \bibfield  {author} {\bibinfo {author} {\bibfnamefont {Tejinder~P.}\
  \bibnamefont {Singh}},\ }\bibfield  {title} {\enquote {\bibinfo {title}
  {Nature does not play dice on the {Planck} scale},}\ }\href@noop {}
  {\bibfield  {journal} {\bibinfo  {journal} {Int. J. Mod. Phys. D}\ }\textbf
  {\bibinfo {volume} {arXiv:2005.06427}},\ \bibinfo {pages}
  {https://doi.org/10.1142/S0218271820430129} (\bibinfo {year}
  {2020}{\natexlab{a}})}\BibitemShut {NoStop}%
\bibitem [{\citenamefont {Singh}(2020{\natexlab{b}})}]{Singhspin}%
  \BibitemOpen
  \bibfield  {author} {\bibinfo {author} {\bibfnamefont {Tejinder~P.}\
  \bibnamefont {Singh}},\ }\bibfield  {title} {\enquote {\bibinfo {title}
  {Octonions, trace dynamics and non-commutative geometry: a case for
  unification in spontaneous quantum gravity},}\ }\href@noop {} {\bibfield
  {journal} {\bibinfo  {journal} {Zeitschrift f\"ur Naturforschung A [to
  appear]}\ }\textbf {\bibinfo {volume} {arXiv:2006.16274v2}} (\bibinfo {year}
  {2020}{\natexlab{b}})}\BibitemShut {NoStop}%
\bibitem [{\citenamefont {S}\ \emph {et~al.}(2020)\citenamefont {S},
  \citenamefont {Pandey},\ and\ \citenamefont {Singh}}]{MPSingh}%
  \BibitemOpen
  \bibfield  {author} {\bibinfo {author} {\bibfnamefont {Meghraj~M}\
  \bibnamefont {S}}, \bibinfo {author} {\bibfnamefont {Abhishek}\ \bibnamefont
  {Pandey}}, \ and\ \bibinfo {author} {\bibfnamefont {Tejinder~P.}\
  \bibnamefont {Singh}},\ }\bibfield  {title} {\enquote {\bibinfo {title} {Why
  does the {Kerr-Newman} black hole have the same gyromagnetic ratio as the
  electron?}}\ }\href@noop {} {\bibfield  {journal} {\bibinfo  {journal}
  {submitted for publication}\ }\textbf {\bibinfo {volume} {arXiv:2006.05392}}
  (\bibinfo {year} {2020})}\BibitemShut {NoStop}%
\bibitem [{\citenamefont {Singh}(2020{\natexlab{c}})}]{Singh2020DA}%
  \BibitemOpen
  \bibfield  {author} {\bibinfo {author} {\bibfnamefont {Tejinder~P.}\
  \bibnamefont {Singh}},\ }\bibfield  {title} {\enquote {\bibinfo {title}
  {Trace dynamics and division algebras: towards quantum gravity and
  unification.}}\ }\href@noop {} {\bibfield  {journal} {\bibinfo  {journal}
  {Zeitschrift f\"ur Naturforschung A [to appear]}\ }\textbf {\bibinfo {volume}
  {arXiv:2009.05574v1 [hep-th]}} (\bibinfo {year}
  {2020}{\natexlab{c}})}\BibitemShut {NoStop}%
\bibitem [{\citenamefont {Palemkota}\ and\ \citenamefont {Singh}(2019 submitted
  for publication)}]{maithresh2019b}%
  \BibitemOpen
  \bibfield  {author} {\bibinfo {author} {\bibfnamefont {Maithresh}\
  \bibnamefont {Palemkota}}\ and\ \bibinfo {author} {\bibfnamefont
  {Tejinder~P.}\ \bibnamefont {Singh}},\ }\bibfield  {title} {\enquote
  {\bibinfo {title} {Black hole entropy from trace dynamics and non-commutative
  geometry},}\ }\href@noop {} {\ \textbf {\bibinfo {volume} {arXiv:1909.02434v2
  [gr-qc]}} (\bibinfo {year} {2019 submitted for publication})}\BibitemShut
  {NoStop}%
\bibitem [{\citenamefont {Singh}(2019)}]{Singh2019qg}%
  \BibitemOpen
  \bibfield  {author} {\bibinfo {author} {\bibfnamefont {Tejinder~P.}\
  \bibnamefont {Singh}},\ }\bibfield  {title} {\enquote {\bibinfo {title}
  {Proposal for a new qantum theory of gravity},}\ }\href@noop {} {\bibfield
  {journal} {\bibinfo  {journal} {Zeitschrift f\"ur Naturforschung A}\ }\textbf
  {\bibinfo {volume} {74}},\ \bibinfo {pages} {617, DOI:
  https://doi.org/10.1515--zna--0079, arXiv:1903.05402} (\bibinfo {year}
  {2019})}\BibitemShut {NoStop}%
\bibitem [{\citenamefont {Vermeulen}\ \emph {et~al.}(2020)\citenamefont
  {Vermeulen}, \citenamefont {Aiello}, \citenamefont {Ejlli}, \citenamefont
  {Griffiths}, \citenamefont {James}, \citenamefont {Dooley},\ and\
  \citenamefont {Grote}}]{s2020experiment}%
  \BibitemOpen
  \bibfield  {author} {\bibinfo {author} {\bibfnamefont {Sander}\ \bibnamefont
  {Vermeulen}}, \bibinfo {author} {\bibfnamefont {Lorenzo}\ \bibnamefont
  {Aiello}}, \bibinfo {author} {\bibfnamefont {Aldo}\ \bibnamefont {Ejlli}},
  \bibinfo {author} {\bibfnamefont {William}\ \bibnamefont {Griffiths}},
  \bibinfo {author} {\bibfnamefont {Alasdair}\ \bibnamefont {James}}, \bibinfo
  {author} {\bibfnamefont {Katherine}\ \bibnamefont {Dooley}}, \ and\ \bibinfo
  {author} {\bibfnamefont {Hartmut}\ \bibnamefont {Grote}},\ }\bibfield
  {title} {\enquote {\bibinfo {title} {An experiment for observing quantum
  gravity phenomena using twin table-top 3d interferometers},}\ }\href@noop {}
  {\bibfield  {journal} {\bibinfo  {journal} {arXiv:2008.04957}\ } (\bibinfo
  {year} {2020})},\ \Eprint {http://arxiv.org/abs/2008.04957} {arXiv:2008.04957
  [gr-qc]} \BibitemShut {NoStop}%
\bibitem [{\citenamefont {Singh}(2020{\natexlab{d}})}]{singh2019qf}%
  \BibitemOpen
  \bibfield  {author} {\bibinfo {author} {\bibfnamefont {Tejinder~P.}\
  \bibnamefont {Singh}},\ }\bibfield  {title} {\enquote {\bibinfo {title} {From
  quantum foundations to spontaneous quantum gravity: an overview of the new
  theory},}\ }\href@noop {} {\bibfield  {journal} {\bibinfo  {journal}
  {Zeitschrift f\"ur Naturforschung A}\ }\textbf {\bibinfo {volume}
  {arXiv:1909.06340 [gr-qc]}},\ \bibinfo {pages} {DOI:
  https://doi.org/10.1515/zna--2020--0073} (\bibinfo {year}
  {2020}{\natexlab{d}})}\BibitemShut {NoStop}%
\end{thebibliography}%

\end{document}